\documentclass[11pt]{article}

\usepackage{fullpage}
\usepackage{CJKutf8} 
\usepackage[utf8]{inputenc}
\usepackage{setspace}
\usepackage{parskip}
\usepackage{titlesec}
\usepackage[section]{placeins}
\usepackage{xcolor}
\usepackage{breakcites}
\usepackage{lineno}
\usepackage{hyphenat}

\usepackage[labelfont=bf]{caption}

\usepackage{times}

\PassOptionsToPackage{hyphens}{url}
\usepackage[colorlinks = true,
            linkcolor = blue,
            urlcolor  = blue,
            citecolor = blue,
            anchorcolor = blue]{hyperref}
\usepackage{etoolbox}


\usepackage{natbib}

\titlespacing{\section}{0pt}{*3}{*1}
\titlespacing{\subsection}{0pt}{*2}{*0.5}
\titlespacing{\subsubsection}{0pt}{*1.5}{0pt}

\usepackage{graphicx}
\usepackage[space]{grffile}
\usepackage{latexsym}
\usepackage{textcomp}
\usepackage{longtable}
\usepackage{tabulary}
\usepackage{booktabs,array,multirow}
\usepackage{amsfonts,amsmath,amssymb}
\providecommand\citet{\cite}
\providecommand\citep{\cite}

\newif\iflatexml\latexmlfalse

\AtBeginDocument{\DeclareGraphicsExtensions{.pdf,.PDF,.eps,.EPS,.png,.PNG,.tif,.TIF,.jpg,.JPG,.jpeg,.JPEG}}

\usepackage[utf8]{inputenc}
\usepackage[english]{babel}

\begin{document}
\begin{CJK}{UTF8}{gbsn}

\title{Early psychosis shows deviations in scaling behaviour within a critical regime}

\author{
Irem Topal\(^1\), Paola Moreno Ancalmo\(^1\), Guillermo Montaña-Valverde\(^1\),\\
Philipp Homan\(^{2,3}\), Wolfram Hinzen\(^{1,4}\)\\[0.5em]
\(^1\) Department of Translation \& Language Sciences, Universitat Pompeu Fabra, Barcelona, Spain\\
\(^2\) Department of Adult Psychiatry and Psychotherapy, University of Zurich, Zurich, Switzerland\\
\(^3\) Neuroscience Center Zurich, University of Zurich and ETH Zurich, Zurich, Switzerland\\
\(^4\) Institut Català de Recerca i Estudis Avançats (ICREA), Barcelona, Spain
}

\date{}

\maketitle

\section*{Abstract}

Accumulating evidence suggests that large-scale brain activity exhibits scale-invariant dynamics consistent with operation in a near-critical regime. Such dynamics have been associated with long-range correlations, efficient information processing, and the emergence of collective organization. While altered criticality-related measures have been reported in psychiatric disorders, previous findings remain fragmented across observables and modalities, making it unclear whether different scaling measures capture a common alteration of large-scale brain dynamics. Here, we investigated scaling properties in resting-state fMRI data from 77 human participants (27 female and 50 male), comprising individuals with early psychosis and healthy controls. We combined a phenomenological renormalization group (PRG) framework with power spectral density (PSD) and detrended fluctuation analysis (DFA) to characterize collective dynamics across scales. In healthy controls, resting-state activity exhibited nontrivial scaling behavior consistent with critical-like organization. Early psychosis participants showed non-trivial scaling behavior, but with systematic shifts in scaling exponents across multiple observables. These findings indicate a reorganization of collective dynamics in early psychosis, with spatial and temporal scaling properties showing distinct alterations that may reflect different changes in proximity to criticality. More broadly, our results suggest that combining coarse-graining approaches with temporal scaling analyses provides a principled framework for studying large-scale brain dynamics in psychiatric disorders.

\section*{Significant Statement}

Resting-state activity in early psychosis preserved broad signatures of critical-like organization, indicating that large-scale collective dynamics remain within a scale-invariant regime despite the presence of clinical pathology. However, patients exhibited systematic and reproducible shifts in scaling exponents across multiple observables, including phenomenological renormalization group (PRG), detrended fluctuation analysis (DFA), and power spectral density (PSD) measures. The consistent pattern of reduced collective integration together with enhanced temporal persistence suggests a reorganization of large-scale brain dynamics rather than a simple breakdown of criticality. These findings further suggest that criticality-inspired scaling frameworks may be sufficiently sensitive to detect subtle alterations in collective brain dynamics even at relatively early stages of psychosis, and identify PRG as a promising approach for studying multiscale brain organization in clinical populations.

\section*{Introduction}

Understanding how large-scale brain activity is organized across space and time remains a central challenge in neuroscience. One influential proposal is that neural systems operate near a critical regime, where collective dynamics exhibit scale invariance and long-range spatiotemporal correlations \cite{beggs2008criticality, chialvo2010emergent, munoz2018colloquium}. Experimental and computational studies have suggested that proximity to criticality may support efficient information processing, maximize dynamic range, and support flexible yet coordinated large-scale neural dynamics \cite{kinouchi2006optimal,shew2013functional}. Consequently, criticality-related scaling properties have increasingly been investigated as potential markers of large-scale brain dynamics underlying both cognition and pathological conditions.

A growing body of work has reported signatures of critical-like dynamics in neural activity using a range of experimental modalities and a variety of computational and statistical approaches. Neuronal avalanche analyses have revealed scale-invariant cascades of activity across species and recording scales \cite{beggs2003neuronal, ponce2018whole, burrows2023microscale, shriki2013neuronal, tagliazucchi-2012}. Long-range temporal correlations have been reported in EEG, MEG, and fMRI signals \cite{linkenkaer2001long, thurner2003scaling, LOMBARDI2021657}, similarly scale-free power spectra have been observed in both electrophysiological \cite{dehghani2010comparative} and BOLD activity \cite{expert2011self, he2011scale}. Together, these findings suggest that spontaneous brain activity may exhibit collective dynamics consistent with near-critical organization.

Criticality-based approaches have increasingly been applied to psychiatric disorders, particularly schizophrenia and psychosis spectrum conditions \cite{zimmern2020brain, hengen2025criticality}. Altered scale-free dynamics have been reported using power spectral density \cite{lee2021alteration}, detrended fluctuation analysis \cite{alamian2022altered}, neuronal avalanche statistics \cite{fekete2021multiscale}, and branching-process measures across fMRI, EEG, and MEG datasets. However, the literature remains methodologically fragmented. Different studies rely on distinct observables, modalities, and preprocessing pipelines, making it difficult to determine whether reported abnormalities reflect a common alteration of scaling organization or instead capture unrelated aspects of neural dynamics. Moreover, many studies focus on single measures in isolation, limiting our understanding of how spatial and temporal scaling properties relate to one another in psychiatric conditions.

A promising framework for addressing these limitations is provided by the phenomenological renormalization group (PRG) approach \cite{PRG-prl, meshulam_coarse--graining_2018}. Inspired by renormalization group methods in statistical physics, PRG method iteratively combines maximally correlated variables to characterize how collective properties evolve across observational scales. PRG-based analyses have been applied across multiple neural recording modalities, including electrophysiology \cite{morales2023quasiuniversal}, MEG \cite{topal2026scaling}, and theoretical models \cite{nicoletti2020scaling}, revealing robust scaling structure in collective dynamics. Recent fMRI studies have shown that coarse-grained neural activity can exhibit nontrivial scaling relations in static observables \cite{ponce2018whole, castro2025interdependent}. Importantly, PRG provides a multiscale framework capable of probing collective organization without requiring explicit assumptions about local network topology or microscopic mechanisms.

Despite growing interest in criticality-inspired analyses, PRG approach has not yet been applied to clinical populations, and its relationship to more established temporal scaling measures such as PSD and DFA remains unclear. This issue is particularly relevant for fMRI, where BOLD signals provide only an indirect measure of neural activity and where the identification of discrete neural events depends strongly on representational choices such as binarization schemes. A key open question is therefore whether different scaling observables converge on a coherent characterization of altered collective dynamics in psychiatric disorders.

In the present study, we addressed this question by complementing PRG analysis with PSD and DFA in resting-state fMRI data from individuals with early psychosis and healthy controls. We first asked whether resting-state fMRI activity exhibits robust spatio-temporal scaling behavior under coarse-graining. We then tested whether scaling exponents differ systematically between groups and whether these effects remain stable across multiple binarization procedures. Across both controls and patients, we observed non-trivial scaling behavior, but early psychosis was associated with systematic shifts in both observables under coarse-graining and temporal scaling properties, suggesting altered multiscale organization of collective brain dynamics. Together, these findings support the use of criticality measures as tools for characterizing large-scale brain organization in psychiatric disorders.

\section*{Materials and Methods}

\subsection*{Dataset}

\subsubsection*{Participants}

Data was obtained from the Human Connectome Project for Early Psychosis (HCP-EP; PI: Dr. Martha Shenton), which recruited 303 participants aged 16 to 35 years with affective psychosis (n = 75), non-affective psychosis (n = 148), and healthy controls (n = 80). Participants with early psychosis were within 5 years of illness onset (mean duration = 1.9 years, standard deviation = 1.4 years) who were diagnosed according to the DSM-V (American Psychiatric Association, 2020). Out of the 303 participants in the original HCP-EP cohort, 100 individuals from the Indiana University site were initially selected to avoid potential confounds related to multi-site data acquisition. This site includes 26 healthy controls (HC) and 74 individuals with non-affective psychosis (schizophrenia, schizophreniform disorder, schizoaffective disorder, delusional disorder, or brief psychotic disorder), as determined by the SCID-5-RV for DSM-5 interview \cite{first2015structured}. All procedures were approved by the Partners Healthcare Human Research Committee/IRB and comply with the regulations set forth by the Declaration of Helsinki. Following quality-control procedures, including preprocessing failures, poor fMRI acquisition quality, insufficient data, and short scan duration, the final sample consisted of 25 healthy controls (11 female, 14 male) and 52 individuals with non-affective psychosis (16 female, 36 male) (see Table~S1). The same quality-controlled sample was used for all PRG, DFA, and PSD analyses.

\subsubsection*{Neuroimaging Acquisition}
The imaging protocol used for HCP-EP included structural MRI, diffusion MRI, and resting-state functional MRI (rs-fMRI) collected during one imaging session. All participants were scanned using the same sequence on a 32-channel head coil Siemens MAGNETOM Prisma 3T scanners at the Indiana University. Prior to scanning, participants completed a 20-minute MRI safety screening in which procedures were described. Participants were instructed to remain still during scanning and a deformable foam cushion was used to minimize head motion. Noise-attenuating headphones and ear stopples were used, which provided excellent noise reduction while permitting adequate auditory perception. Trained study staff reviewed all images in real-time on the scanning console for quality assurance. If there was a detectable problem, the scan was repeated. Blood was drawn during the MRI visit prior to the scan. The specific imaging protocol took approximately 65 minutes to complete and included a localizer and autoalign scout, and structural T1w (MPRAGE; 0.8 mm isotropic; Repetition time [TR] 2400 ms; Inversion time 1000 ms; Flip angle 8°) and T2w (SPACE; 0.8 mm isotropic; TR 3200 ms; Echo time [TE] 563 ms) scans. Participants underwent about 23
minutes of rs-fMR (2 mm isotropic; multiband (MB) acceleration × 8; TR  800 ms) acquired across four 5-minute 46 seconds scans (420 measurements in each), and in 2 blocks, each block with one scan in Anterior-Posterior (AP) and 1 in Posterior Anterior (PA) phase encoding. During the rs-fMRI scans, participants were asked to keep their eyes open and focus their gaze on a black cross against a light gray background. Detailed imaging protocols and scanning parameters can be found in HCP (\url{https://www.humanconnectome.org/study/human-connectome-project-for-early-psychosis}). 

\subsubsection*{fMRI Preprocessing}
The fMRI dataset was organised in the Brain Imaging Data Structure (BIDS) standard so as to be passed to fMRIPrep (version 24.1) \cite{esteban2019fmriprep}. Detailed steps and functions are provided in the Supplementary Material (see Image Pre-processing with fMRIPrep).

\subsubsection*{Regressing Confounds and Brain Parcellation}
To reduce potential confounding effects in the fMRI data, several regressors were included in analyses. First, we applied a high-pass filter with a cutoff frequency of $0.008$ Hz to remove low-frequency signals that can be introduced by physiological and scanner noise sources. Next, to reduce the effects of head motion, we regressed out the six rigid-body motion parameters, which have been previously demonstrated to introduce bias in group comparisons \cite{power2012spurious}. These motion parameters include the transition on the three axes (x, y, z) and the respective rotation (α,β, γ), which are estimated relative to a reference image \cite{friston1996movement}. Additionally, to minimize the impact of non-neuronal BOLD signal fluctuations, we regressed out signals from white matter and cerebrospinal fluid \cite{fox2005human}. These signals, as well as the motion parameters, were
expanded using the first temporal derivatives and their quadratic terms to capture potential non-linear effects of these noise sources \cite{satterthwaite2013improved}. These regression steps were based on a denoising strategy proposed by \cite{wang2024continuous}, which aims to improve the quality of fMRI connectivity studies. This regression procedure was applied to the corresponding preprocessed NifTI images through a brain atlas for denoising. Although there is no consensus on the optimal spatial brain parcellation \cite{eickhoff2018imaging}, all neuroimaging data in this study were preprocessed using the Schaefer parcellation \cite{schaefer2018local}, and we specifically employed the 1000-region resolution. This relatively fine-grained parcellation is well suited to our PRG framework, as it enables the exploration of scaling behavior by maximizing the number of degrees of freedom available for coarse-graining across scales. As a robustness analysis, we repeated the complete PRG analysis using
progressively coarser Schaefer parcellations of 800, 600, and 400 ROIs. This analysis was performed to assess the sensitivity of the results to the high dimensionality of the correlation estimation relative to the length of the fMRI time series.

\subsubsection*{Global Signal Extraction}

For the global-signal analyses, we used the same fMRIPrep preprocessed BOLD images in MNI152NLin2009cAsym space described above. In contrast to the parcel-based PRG analyses, voxelwise time series were extracted directly from the whole-brain mask using \texttt{NiftiMasker} from Nilearn. Spatial smoothing was applied during extraction using a Gaussian kernel with 6 mm full-width at half-maximum (FWHM). To reduce the influence of non-neuronal fluctuations, voxelwise time series were denoised using a nuisance-regression procedure including high-pass filtering, rigid-body motion regressors, and white-matter/cerebrospinal-fluid signals, using the full motion and WM/CSF models implemented in Nilearn. Confounds were standardized prior to regression, while voxelwise signals were not detrended or standardized at this stage. The denoised global signal for each subject was then computed as the mean of the denoised voxelwise time series across all voxels within the brain mask.

\subsection*{Pre-processing and neural event identification}

To extract discrete event sequences from continuous BOLD signals, we applied a binarization procedure at the level of individual regions of interest (ROIs). For each scan, ROI time series were first z-scored, yielding normalized signals $z_i(t)$ for $1 < i < N$, where $N$ denotes the number of ROIs. Binarization was then performed using a fixed threshold $\sigma$, expressed in units of standard deviation. All analyses were repeated across multiple thresholds to assess robustness. Events were defined as all time points at which the signal exceeded the threshold ($z_i(t) > \sigma$). This method captures the full duration of suprathreshold activity, thereby preserving information about sustained high-amplitude fluctuations in the BOLD dynamics.

To further assess the robustness of our results with respect to the choice of binarization scheme, we additionally considered two alternative definitions of events. First, following point-process approaches to fMRI analysis \cite{tagliazucchi-2012}, events were defined as upward threshold crossings, such that a binary event was assigned at time $t$ when $z_i(t) > \sigma$ and $z_i(t-1) \leq \sigma$. This approach emphasizes the onset of high-amplitude fluctuations and results in a sparse point-process representation of the signal.

Second, events were defined as local extrema within suprathreshold excursions \cite{genetic-criticality}. In this case, a binary event was assigned at time $t$ if the signal exhibited a local maximum within positive excursions above $\sigma$, or a local minimum within negative excursions below $-\sigma$. This yields a sparse set of events corresponding to peaks of large-amplitude fluctuations. A comparison of binarization schemes is provided in the Supplementary Information Fig.~S1.

\subsection*{Phenomenological Renormalization Group}

Coarse-graining aims to simplify high-dimensional systems in order to characterize their collective behavior. This concept, originating from statistical physics and applied to neuronal activity \cite{PRG-prl}, involves tracking the joint probability distribution of microscopic variables as the observation scale is changed. In the current fMRI application, microscopic variables are the regional BOLD signals.
The approach is based on clustering pairs of the most correlated regions by summing up their activities. First, the entries of correlation coefficient matrix $c_{ij}$ are calculated by:
\begin{equation}
    c_{ij} = \frac{\langle \delta x_i \delta x_j \rangle} { [\langle \delta x_i^2 \rangle \langle \delta x_j^2 \rangle]^{1/2} }
\end{equation}
where $\delta x_i = x - \langle x_i \rangle $. We search for the most strongly correlated pair $(i,j)$ by finding the largest non-diagonal element in the matrix and sum these regions' activities to define the coarse-grained variable as:
\begin{equation}
    x_i^{(2)} = Z_i^{(2)} (x_i + x_j)
\end{equation}
where $Z_i^{(2)}$ normalizes the coarse-grained variable. We normalize each signal such that the average amplitude of the nonzero values equals one, to eliminate arbitrariness while maintaining the significance of nonzero activity. We apply this normalization at every step of coarse-graining. Each region can only participate in one pair so we remove the pair $(i,j)$, then continue by selecting the next most strongly correlated pair, until the original $N$ variables have been grouped into $N/2$ pairs. For the next step of the procedure, we calculate new correlations between pairs of \emph{coarse-grained variables} and then cluster the maximally correlated pairs by summing up their activity and iterate this until all pairs are grouped. At each iteration, coarse-grained variables (represented by $x_i^{(K)}$) are clusters of size $K=2^k$, where $k$ is the PRG step. 

PRG allows the detection of nontrivial scale-invariant behavior by tracking how collective properties evolve as the observation scale increases during coarse-graining.

A central insight of the renormalization group is the existence of non-Gaussian fixed points, where the summation of correlated variables often leads to non-Gaussian distributions at intermediate scales, in contrast to the Gaussian behavior observed for independent variables. Thus, the first observable we check is the distribution of normalized activity. 

The second observable is the log-probability of silence of the coarse-grained variables, $\ln P_0$. We examined its scaling with cluster size $K$, following the relation $-\ln P_0 \propto K^\beta$. At each PRG step, silence probabilities were computed for individual coarse-grained variables and averaged across clusters of corresponding size $K$.

A third static quantity we calculate is the variance of non-normalized coarse-grained variables at each scale. 
\begin{equation}
    Var (K) = \langle (x_i^{(K)})^2 \rangle - \langle x_i^{(K)} \rangle ^2
\end{equation}
The variance of the coarse-grained variables scales with cluster size according to a power-law, $Var(K) = K^\alpha$ where the exponent reflects the degree of correlation in the system. If the independent variables are coarse-grained, the variance grows linearly with cluster size ($Var(K) = K^1$), while in fully correlated systems, it grows quadratically ($Var(K) = K^2$). Observing a scaling with $1< \alpha < 2$ indicates nontrivial scaling behavior, suggesting that the system exhibits intermediate correlations characterized by an underlying self-similar, coordinated structure.

\begin{figure}[h!]
\begin{center}
\includegraphics[width=\columnwidth]{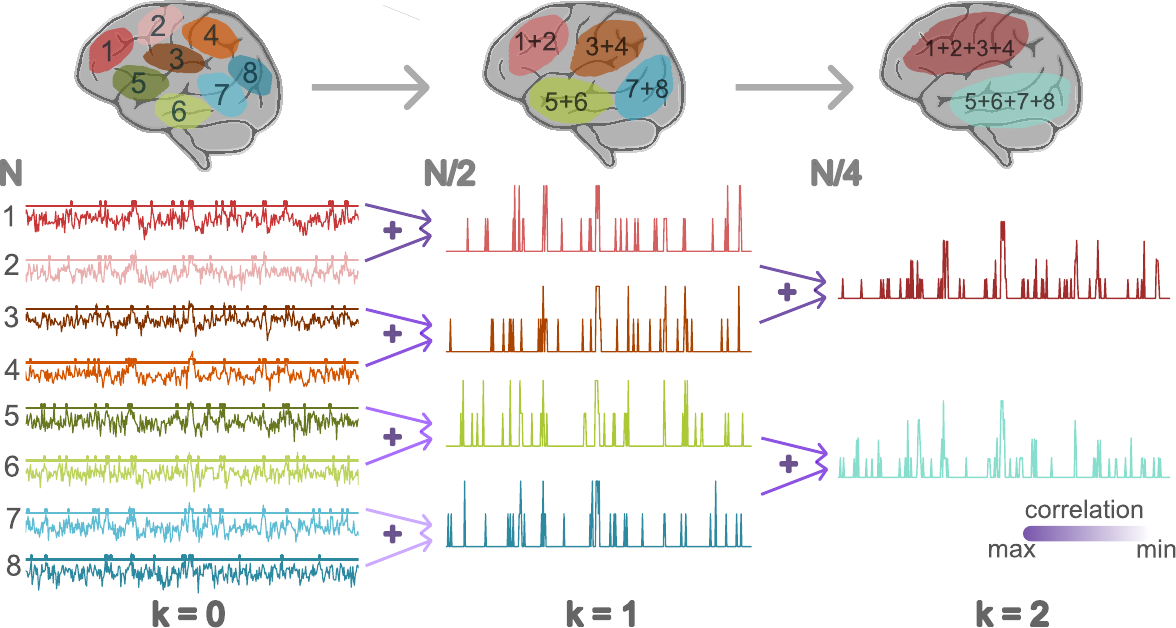}
\caption{\textbf{Schematic illustration of the PRG procedure.} Initially ($k=0$), each variable is binarized independently. At each coarse-graining step $k$, pairs of maximally correlated variables are summed, generating clusters of size $K = 2^k$ and reducing the total number of clusters to $N_K = N/2^k$. Scaling properties are then evaluated as a function of the coarse-graining scale.}
\label{prg-scheme}
\end{center}
\end{figure}

The last observable is the eigenspectrum of covariance matrices computed within clusters of size $K$. If correlations are self-similar across scales, this structure should also be reflected within individual clusters generated during coarse-graining. For each cluster, we therefore computed the covariance matrix using the original regional activity of the regions belonging to that cluster. Importantly, although the covariance matrices are calculated from the original regional signals, all other observables are computed using the coarse-grained variables. The eigenvalues $\lambda$ of each covariance matrix were ordered from largest to smallest, and the eigenspectra were subsequently averaged across all clusters at a given coarse-graining level to obtain the representative spectrum for scale $K$. The eigenspectrum of in-cluster covariance matrices scales as:

\begin{equation}
    \lambda \propto (\frac{rank}{K})^{- \mu}
\end{equation}
The eigenspectrum exhibits scaling in two distinct ways: First, when the rank is normalized by $K$, the spectra corresponding to clusters of varying sizes collapses. Second, the eigenvalues display a power-law decay with respect to rank, which is then followed by an exponential cutoff due to the limitations imposed by the finite size of the system.

The last scaling relation we calculated arises from the dynamical scaling of the coarse-grained variables, where we measure the autocorrelation function of an individual cluster $i$ of size $K$:
\begin{equation}
    C_i = \frac{\langle \delta x_i^{(K)}(t_0) \delta x_i^{(K)}(t_0 + t) \rangle} {Var(x_i^{(K)})}
\end{equation}
and the average across all clusters:
\begin{equation}
    C_K = \frac{1}{N_K} \sum_i^{N_K} C_i
\end{equation}
where $N_K$ is the number of clusters of size $K$. We determine the characteristic time ($\tau_c(K)$) of mean-normalized autocorrelation functions at each scale by fitting an exponential decay. Under dynamical scaling, this characteristic time is expected to scale with cluster size, $\tau_c(K) = K^z$. When coarse-graining two signals that are either perfectly correlated with identical timescales or completely uncorrelated, the resulting scaling exponent $z$ is zero, indicating no dynamical scaling. However, a scaling exponent $z>0$ reveals the emergence of nontrivial temporal scaling through the coarse-graining process.

\subsection*{Detrended Fluctuation Analysis}
 
Detrended Fluctuation Analysis (DFA) provides a robust framework for quantifying scale-free temporal structure and long-range temporal correlations in nonstationary biological signals \cite{peng1994dfa, hardstone2012detrended}. Long-range temporal correlations have been proposed to reflect adaptive multiscale organization in physiological systems, while alterations in these scaling properties have been associated with aging and disease \cite{goldberger2002fractal}. Here, DFA was applied to the denoised global signal to quantify long-range temporal correlations and assess potential alterations in temporal scaling organization associated with early psychosis. The integrated signal profile was first obtained by computing the cumulative sum of the mean-subtracted time series. The profile was subsequently divided into non-overlapping windows of varying size $n$, with scales ranging from a minimum window of $3$ timepoints to a maximum of one quarter of the total signal length, distributed across $16$ logarithmically spaced values. To improve estimation stability, this procedure was additionally applied to the time-reversed profile, and all resulting segments were pooled. For each window, a linear trend was estimated and removed, and the root mean square of the residuals was computed to yield the fluctuation function $F(n)$. The DFA scaling exponent $\alpha_{DFA}$ was estimated by linear regression of $\log F(n)$ against $\log n$ over the range $n=4$--$100$, corresponding to the range over which the fluctuation function exhibited approximately linear scaling in log--log space. An exponent of $\alpha_{DFA}=0.5$ indicates temporally uncorrelated noise, values of $\alpha_{DFA} > 0.5$ indicate the presence of long-range temporal correlations, and $\alpha_{DFA} \approx 1$ is characteristic of scale-free, $1/f$ dynamics.

\subsection*{Power Spectral Density Analysis}

Power spectral analysis characterizes how signal power is distributed across frequencies. In resting-state fMRI, spontaneous BOLD fluctuations are dominated by low-frequency components ($\lesssim 0.2$ Hz), reflecting the slow dynamics of cerebrovascular responses \cite{Pang_psd}. Several studies have shown that the BOLD power spectrum follows a power-law form, $P(f)\propto f^{-\beta_{PSD}}$ where $P$ is the power and $f$ is the frequency, at low frequencies, indicating scale-free temporal organization. The exponent $\beta_{PSD}$ quantifies temporal correlations in the signal, with larger values associated with stronger long-range memory and smaller values linked to more efficient information processing. Motivated by these observations, we estimated the power spectral density (PSD) of the global fMRI signal to quantify its scale-free temporal dynamics. PSD was estimated using Welch’s method \cite{welch_psd}. The time series were divided into overlapping segments with length equal to one quarter of the total signal duration, with $50\%$ overlap between adjacent segments. A constant detrending procedure was applied prior to spectral estimation, and the sampling frequency was set to $1.25$ Hz. To estimate the scaling exponent both frequency and power were log-transformed, and linear regression was performed in log-log space over the frequency range $0.01{-}0.2$ Hz, corresponding to the frequency range commonly associated with spontaneous resting-state fluctuations \cite{psd_frange}. The resulting slope, $\beta_{PSD}$, was taken as an estimate of the power-law scaling exponent of the global signal.

\subsection*{Null models}
To verify that the observed critical statistics arise from the intrinsic spatiotemporal organization of neuronal activity captured in the BOLD signals, rather than from trivial or random features of the data, we used an appropriate null model. Surrogate BOLD time series were generated by shuffling the phases for each region independently \cite{theiler92}. This procedure preserves both the empirical amplitude distribution and the linear autocorrelation structure (power spectral density) of the original BOLD signals, while effectively disrupting the phase-synchrony across regions.

\subsection*{Power-Law Fitting and Statistical Analysis}
Scaling exponents were estimated using linear least-squares regression in log-log space. For the silence probability, fits were performed over all available cluster sizes. In a small number of subjects, $P_0(K)$ vanished at the largest cluster sizes; these points were excluded from the fit. Variance scaling was fitted over the full range of cluster sizes. Dynamic scaling ($z$) was estimated after excluding the smallest (raw variables) and the largest cluster sizes. For the eigenspectrum, fits were performed over normalized ranks $1/64 \leq r/K < 40/64$ ($K=64$), excluding the dominant first eigenvalue and the low-rank tail where deviations from power-law scaling become apparent. Regression standard errors were computed for each individual exponent estimate. Because the analyzed observables did not correspond to probability distributions, alternative maximum-likelihood procedures commonly used for fitting heavy-tailed distributions were not applied. Group differences in scaling exponents were assessed using Welch’s two-sample \textit{t}-test. Associations between scaling exponents were evaluated using the Pearson correlation coefficient.

\section*{Results}
In our analysis, we used the HCP-EP resting-state dataset. For the PRG analysis, we employed a 1000-region parcellation of the brain. The BOLD time series extracted for each region of interest (ROI) were z-normalized and subsequently binarized using a threshold of $1.5$ SD. This procedure enables the identification of discrete neural events underlying the continuous BOLD signals. The idea is to track multiple static and dynamic quantities across successive PRG steps. We assessed whether scaling behavior under PRG is altered in early psychosis and characterized the nature of these changes. For PSD and DFA analyses, we used the global signal derived from the fMRI data.

\subsection*{Whole-Brain Resting-State Activity Exhibits Critical Scaling in Healthy Population}

We performed PRG analysis by iteratively grouping the most strongly correlated regions at each step, continuing until no further pairs could be formed (see Materials and Methods and Fig.~\ref{prg-scheme} for details).

\begin{figure}[h!]
\begin{center}
\includegraphics[width=\columnwidth]{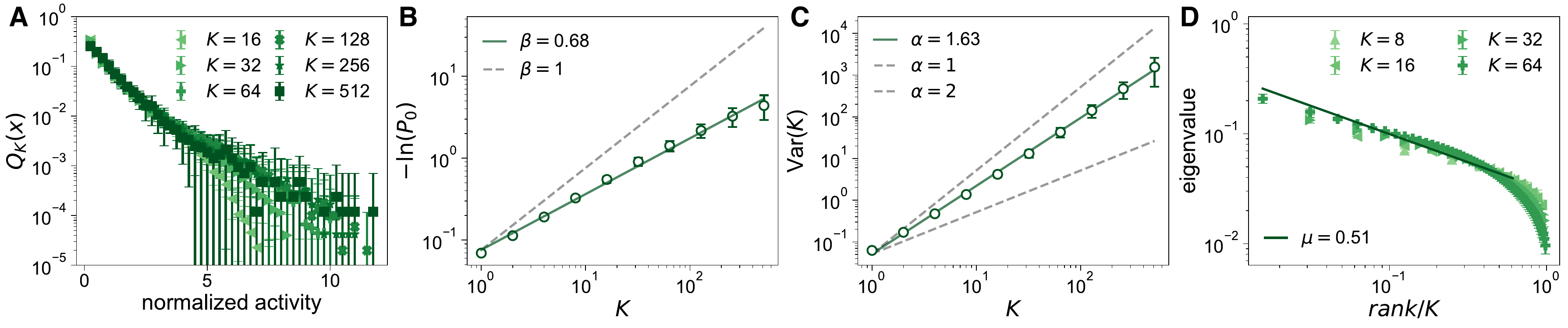}
\caption{\textbf{Resting-state whole-brain activity reveals statistical signatures of criticality.} \textbf{A.} Average probability distribution for normalized nonzero activity across 25 control subjects. Darker green indicates larger cluster sizes. \textbf{B.} Average log-probability of silence in coarse-grained variables ($ln P_0$) as a function of cluster size, $K$. The solid line represents a linear least-squares fit of $-ln P_0 = K^\beta$, where $\beta = 0.68$ ($R^2=0.99$, $p < 10^{-9}$). The gray dashed line indicates predicted behaviour of uncorrelated variables $\beta = 1$. \textbf{C.} Average variance (Var) of the non-normalized coarse-grained variables across subjects as a function of the cluster size, $K$. The solid line corresponds to the linear least-square fit,  $log \ Var(K) = \alpha \cdot logK + b$, where $\alpha = 1.63$ ($R^2=0.99$, $p < 10^{-9}$). Gray dashed lines mark reference scaling behaviors: linear ($\alpha = 1$), corresponding to uncorrelated variables, and quadratic ($\alpha = 2$), corresponding to fully correlated variables. \textbf{D.} Eigenvalues of cluster covariance matrices  for cluster sizes $K \leq 64$. The solid line is a least squares fit to $log \ \lambda = log \ b (rank/K)^{-\mu}$ performed for $K=64$. Error bars represent standard deviation across subjects (SD).}
\label{controls}
\end{center}
\end{figure}

We first asked whether resting-state whole-brain activity in healthy controls displays scaling behaviour under iterative coarse-graining. To this end, we applied PRG procedure to 1000 regions of interest. Fig.~\ref{controls} summarizes the behavior of the coarse-grained activity across 25 healthy subjects. In Fig.~\ref{controls}A, we show the average probability distribution of normalized nonzero activity for different cluster sizes, $Q_K(x)$. As the cluster size increases, the distributions retain a non-Gaussian fixed form, indicating that the statistics of activity remain structured across coarse-graining steps. This persistence of the distribution across scales suggests that a nontrivial fixed point of RG flow.

Then we measured the probability of silence of coarse-grained variables as a function of the cluster size. As shown in Fig.~\ref{controls}B, $-\ln{P_0}$, increases with cluster size and exhibits a clear scaling, with an average exponent $\beta = 0.69 \pm 0.04$ across $25$ healthy subjects. Importantly, this increase is slower than expected for a set of independent variables $\beta=1$, indicating that the observed behavior points to nontrivial statistical dependencies across coarse-graining scales. 

We then examined how fluctuations grow as variables are summed. For this analysis, the variance was computed from the non-normalized coarse-grained variables, so that the growth of fluctuations with cluster size could be directly assessed. As shown in Fig.~\ref{controls}C, we observe a perfect scaling for the variance of the non-normalized variables with an average $\alpha = 1.61 \pm 0.06$ in healthy population. Moreover, the observed scaling lies between the limiting cases expected for uncorrelated ($\alpha=1$) and fully correlated variables ($\alpha=2$). This intermediate behavior indicates that the activity is neither dominated by independent fluctuations nor by complete redundancy, but instead exhibits non-trivial scaling over multiple scales. Our estimates of the variance and silence probability exponents are consistent with previously reported fMRI measurements of PRG scaling behavior \cite{ponce2023critical}. 

As the last static quantity, instead of focusing on the coarse-grained variables themselves, we turn our attention to the internal structure of the clusters obtained at each coarse-graining step by analyzing their covariance eigenspectra. Fig.~\ref{controls}D shows that the eigenvalue spectra for different cluster sizes collapse when plotted against normalized rank, and that they follow a clear power law decay with an average exponent $\mu = 0.51 \pm 0.03$ (n=25). This result indicates that the hierarchy of collective modes is preserved across coarse-graining levels, providing an additional signature of nontrivial scaling organization. Eigenspectra were computed up to $K \leq 64$, excluding the largest clusters to control finite-sample effects. In high-dimensional settings where the number of dimensions, $N$, is not negligible relative to number of samples, $T$, sample covariance eigenvalues are known to be systematically biased and spread due to noise \cite{kong2017spectrum}. Restricting cluster size keeps $q=N/T$ small and improves the reliability of the estimated spectra, which is especially important for short fMRI time series \cite{meshulam_coarse--graining_2018}.

\begin{figure}[h!]
\begin{center}
\includegraphics[width=0.8\columnwidth]{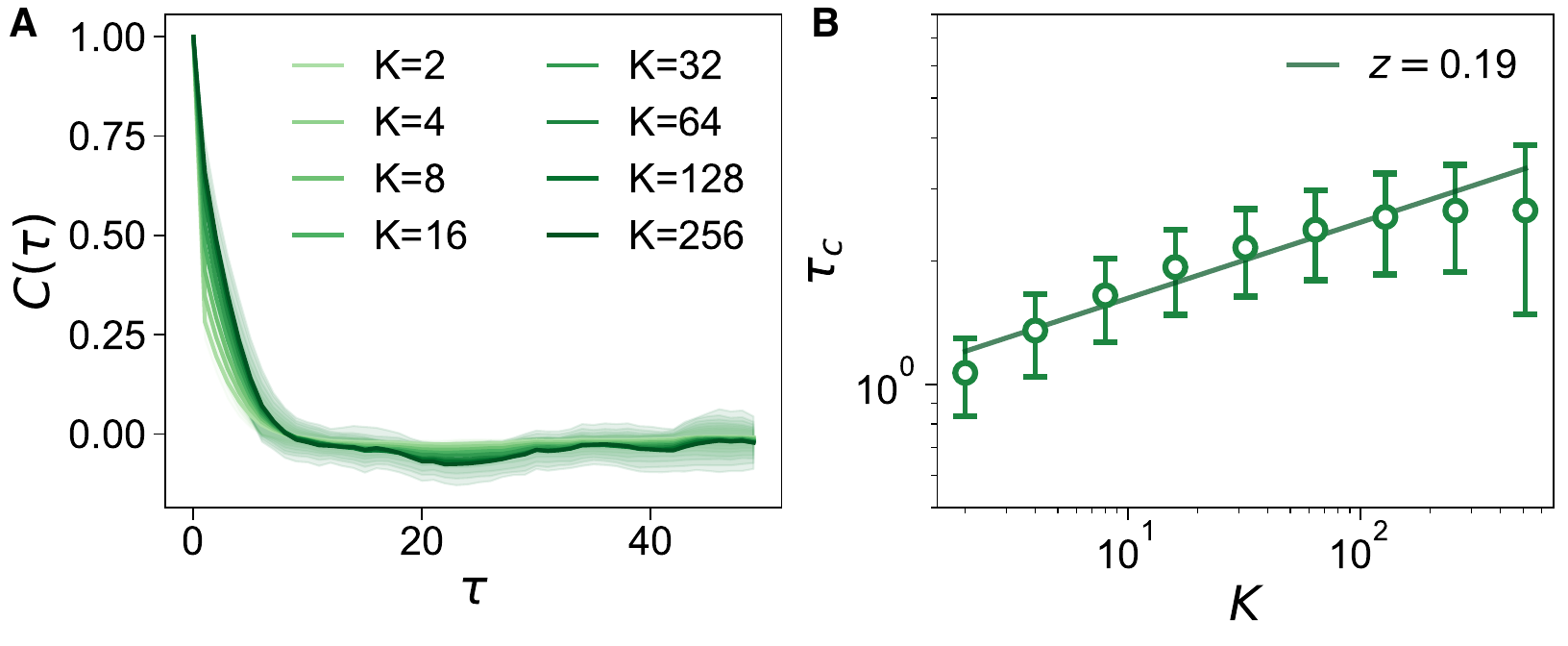}
\caption{\textbf{Dynamic scaling under coarse-graining for healty controls.}
\textbf{A.} Average autocorrelation functions $C(\tau)$ for each cluster size $K$. 
For each subject, autocorrelations are computed and averaged across clusters; 
the curves shown correspond to the mean across subjects. Shaded regions indicate 
the standard deviation. \textbf{B.} Autocorrelation time $\tau_c$, obtained by fitting an exponential decay to the average autocorrelation function of each healthy subject individually. The points show the mean $\tau_c$ across controls as a function of $K$, with error bars denoting the standard deviation. A power-law scaling $\tau_c \propto K^z$ is observed; the reported exponent $z=0.19$ is obtained from a least-squares fit to the averaged $\tau_c$ values ($R^2=0.94$, $p < 10^{-4}$).}
\label{controls-dy}
\end{center}
\end{figure}

We next examined how temporal correlations evolve under coarse-graining. For each level of coarse-graining, we computed the mean autocorrelation function for the corresponding cluster size $K$. As shown in Fig.~\ref{controls-dy}A, the autocorrelation functions decay more slowly as the cluster size increases. We fit an exponential function and find the characteristic time, $\tau_c$. Figure~\ref{controls-dy}B shows that this timescale increases with cluster size. Importantly, this growth follows a clear scaling relationship in double logarithmic scale, indicating that temporal correlations are organized across scales with an average exponent $z=0.17 \pm 0.04$ (n=25).

The relatively small inter-subject variability of the estimated exponents indicates that the observed scaling behavior is robust across healthy subjects. As a control, we found that the nontrivial scaling behavior uncovered by the PRG analysis vanishes across all healthy subjects when the cross-correlation structure is destroyed by phase-shuffling the regional BOLD signals (see Supplementary Fig.~S2).

Taken together, these results show that resting-state whole-brain activity in healthy controls exhibits several statistical signatures expected from a system near criticality. The persistence of the activity distribution across scales, the nontrivial scaling of silence probability, the intermediate scaling of variance, and the scale-free organization of covariance eigenmodes all support the presence of critical-like structure in the resting-state dynamics.

\subsection*{Altered Scaling Exponents in Early Psychosis}

We next asked whether the scaling behavior observed in healthy controls is preserved in early psychosis patients. Applying the same analysis pipeline, we find that resting-state activity in patients also exhibits nontrivial scaling behavior across all measured observables (see Supplementary Fig.~S3), indicating that the overall phenomenology of scale-invariant organization is maintained.

\begin{figure}[h!]
\begin{center}
\includegraphics[width=\columnwidth]{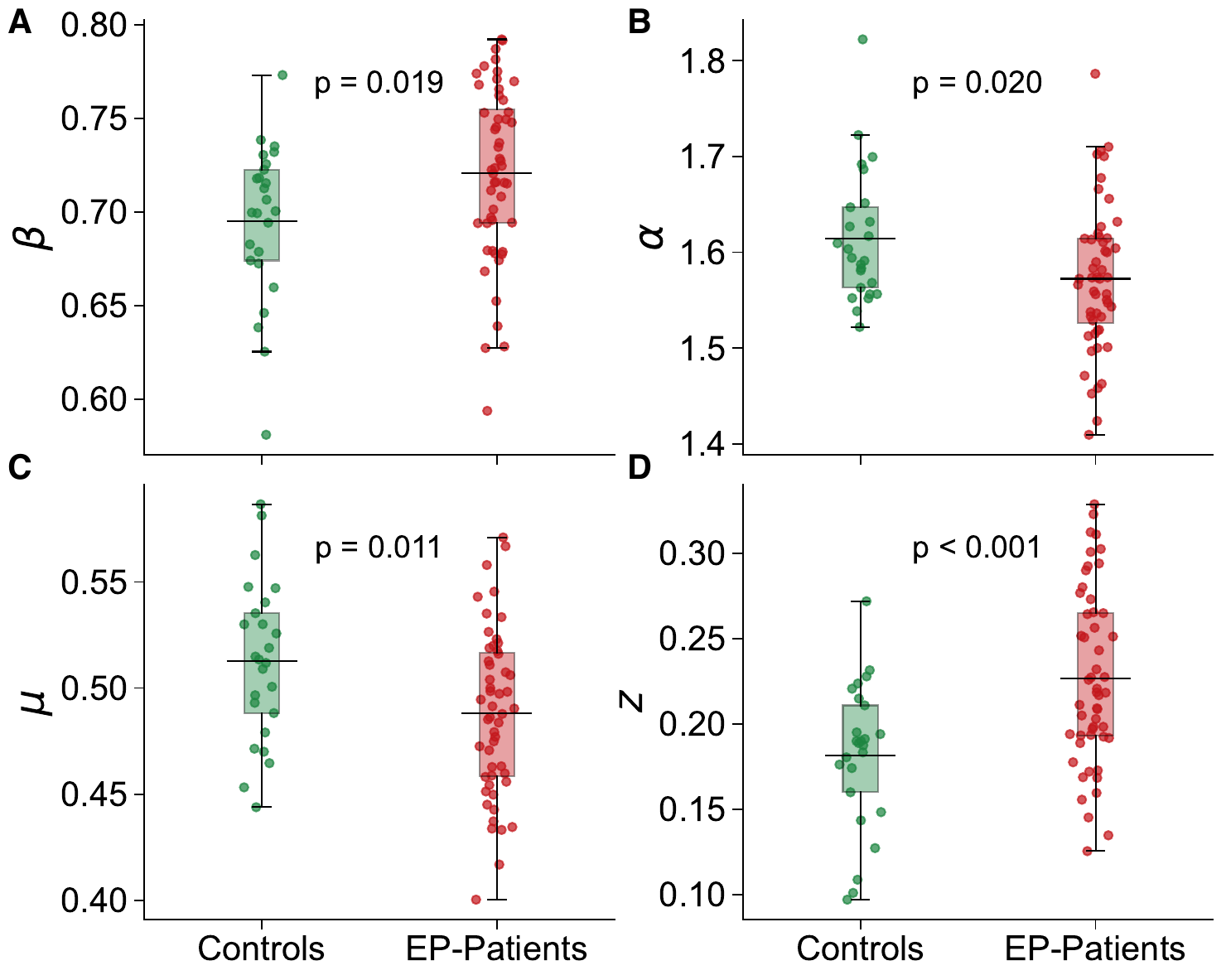}
\caption{\textbf{Altered scaling exponents in early psychosis.} Distributions of scaling exponents across subjects for healthy controls (green) and EP-patients (red). Each panel corresponds to a different exponent obtained from PRG analysis: \textbf{A.} silence probability exponent $\beta$, \textbf{B.} variance scaling exponent $\alpha$, \textbf{C.} eigenspectrum exponent $\mu$, and \textbf{D.} dynamical scaling exponent $z$. For each group, boxplots summarize the distribution of subject-level exponents, with boxes indicating the interquartile range and whiskers extending to the data range excluding outliers (fliers not shown). Horizontal black lines indicate group means. Statistical comparisons between groups were performed using Welch’s t-tests, and corresponding $p$-values are displayed within each panel.}
\label{prg-exponents}
\end{center}
\end{figure}

However, a direct comparison of scaling exponents reveals systematic differences between groups. Fig.~\ref{prg-exponents} shows the distribution of PRG exponents across subjects for both healthy controls and early psychosis patients. To assess the numerical precision of the exponent estimation, we quantified the regression standard error of each individual least-squares fit. Across all four exponents, the mean standard errors were similar between healthy controls and early psychosis patients (HC: 0.014--0.020; EP-Patients: 0.015--0.020), indicating that the uncertainty associated with the fitting procedure was small relative to the observed inter-subject variability. We observe significant group differences in all four scaling measures: the silence probability exponent $\beta$ is increased in patients ($p = 0.019$), while the variance exponent $\alpha$ is reduced ($p = 0.02$). Consistent with theoretical expectations, these exponents exhibit anticorrelated behaviour. These results were robust to the fitting range. Excluding the largest one and two coarse-graining levels, where finite-size effects are expected to be strongest, did not alter the statistical significance of the group differences. Within the PRG framework, the observed incerase in $\beta$ together with the reduction in $\alpha$ suggest a shift in patients toward weaker collective organization under coarse graining. Similarly, the eigenspectrum exponent $\mu$ is decreased ($p = 0.011$), and the dynamical exponent $z$ is increased ($p < 0.001$). A reduction in $\mu$ corresponds to a slower decay of the eigenspectrum, indicating that fluctuations are distributed across a broader range of collective modes rather than being dominated by a small number of large-scale components. In contrast, the increase in $z$ indicates stronger persistence of autocorrelations across coarse graining scales, since $z=0$ corresponds to the case of uncorrelated variables. These results indicate that, although scaling behavior persists in early psychosis, the underlying scaling laws are systematically altered. Rather than reflecting a uniform loss of critical-like structure, the observed changes suggest a differential reorganization of static and dynamical collective properties, consistent with a modification of the effective dynamical regime governing large-scale brain activity. Given the relatively high dimensionality of the 1000-ROI parcellation compared with the length of the BOLD signals, we repeated the complete PRG analysis using progressively coarser parcellations of 800, 600, and 400 ROIs to assess the potential influence of high-dimensional sampling noise on the PRG hierarchy.  Differences in $\alpha$, $\mu$, and $z$ remained significant at all three resolutions, while the difference in $\beta$ was significant at 600 and 400 ROIs (Table~S2-A). Thus, the main pattern of scaling alterations was preserved under substantial reductions in spatial dimensionality.

\begin{figure}[h!]
\begin{center}
\includegraphics[width=\columnwidth]{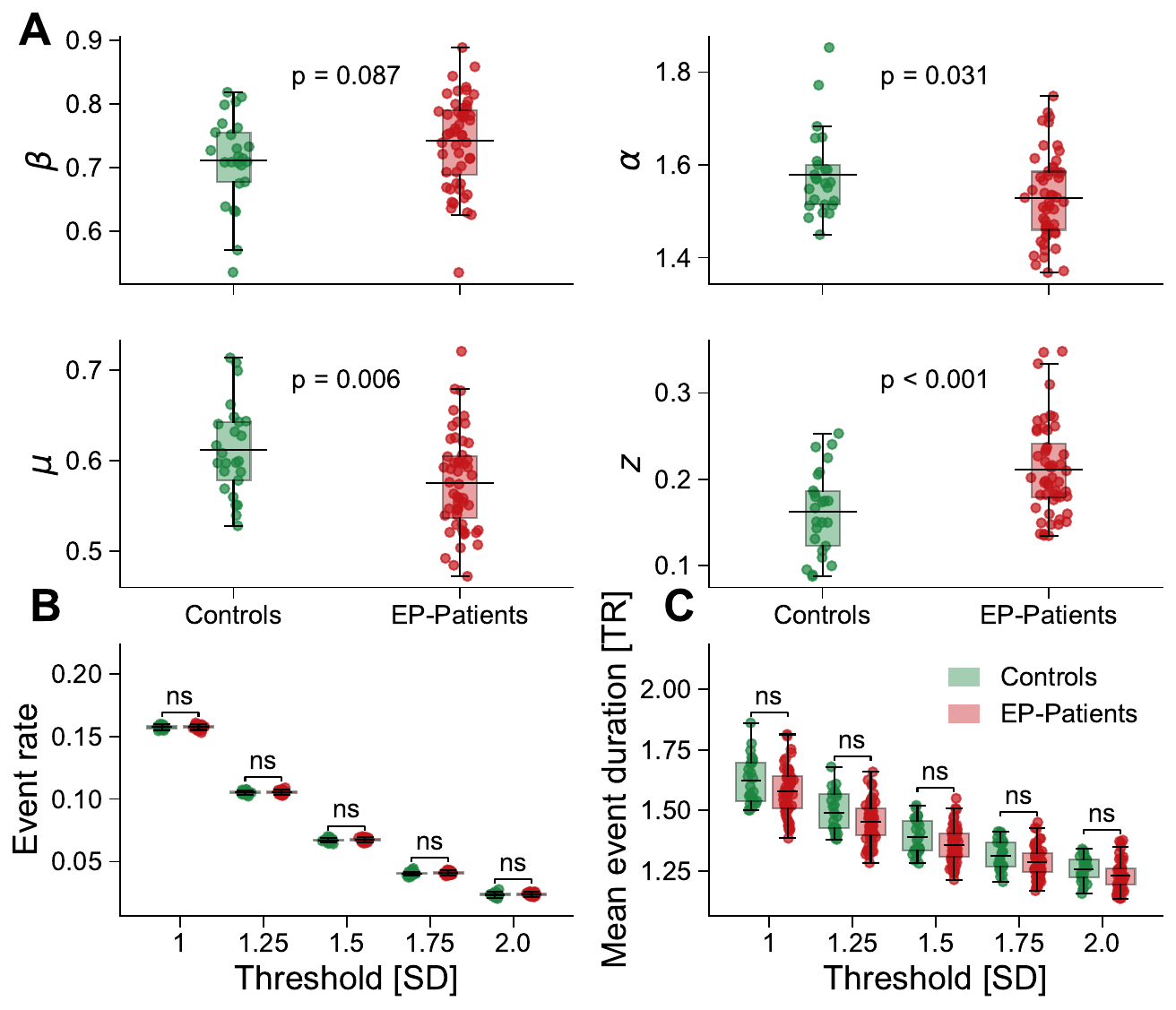}
\caption{\textbf{Robustness analyses of PRG scaling differences between healthy controls and early psychosis patients.} \textbf{A.} Held-out validation of the PRG hierarchy. For each participant, the hierarchical pairing structure was inferred from the first half of the recording and kept fixed while the scaling exponents $\beta$, $\alpha$, $\mu$, and $z$ were estimated from the second half. Group comparisons are shown for each exponent.
\textbf{B.} Event rate across binarization thresholds.
\textbf{C.} Mean duration of suprathreshold events across the same thresholds. Event rate and mean event duration did not differ significantly between groups at any threshold. Group comparisons were performed using Welch's $t$-tests.
}
\label{cross-val}
\end{center}
\end{figure}

We further assessed the robustness of the subject-level scaling exponents using two complementary temporal validation analyses. First, PRG exponents were estimated independently from the first and second halves of each recording. The overall pattern of group differences was reproduced (first half: $\beta$, $p=0.10$; $\alpha$, $p=0.02$; $\mu$, $p=0.01$; $z$, $p=0.008$; second half: $\beta$, $p=0.062$; $\alpha$, $p=0.014$; $\mu$, $p=0.01$; $z$, $p=0.001$). We then performed a more stringent held-out validation in which the PRG hierarchy was inferred exclusively from the first half of each recording and subsequently kept fixed while the scaling exponents were estimated from the second half. Group differences remained significant under this procedure (Fig.~\ref{cross-val}A), indicating that the inferred coarse-graining hierarchy
generalized to unseen data.

To assess the robustness of PRG exponents with respect to binarization choice, we repeated the analysis across thresholds ranging from $1$ to $2$ SD. Within both control and patient groups, all exponents exhibited relatively limited variability across thresholds, indicating that the non-trivial scaling properties are stable with respect to threshold choice. Importantly, the direction of the group differences remained consistent across the entire threshold range for all exponents (see Fig.~S4). In particular, $\alpha$ and $z$ showed significant differences between patients and controls at all tested thresholds (Welch’s t-test, $\alpha$: $p = 0.003{-}0.020$; $z$: $p < 5\times10^{-4}$), with highly stable effect magnitudes. Differences in $\mu$ also remained consistent across thresholds and became progressively stronger at higher thresholds ($p = 0.064$ at $1$ SD, decreasing to $p = 0.003$ at $2$ SD). In contrast, $\beta$ displayed a more pronounced dependence on the binarization threshold, with group differences becoming progressively stronger at higher thresholds ($p = 0.60$ at $1$ SD, decreasing to $p = 0.008$ at $2$ SD). Together, these results indicate that the observed alterations in PRG scaling exponents are not driven by a specific choice of binarization threshold, but instead reflect stable differences between clinical and control populations.

We further examined the robustness of these results with respect to alternative binarization schemes. Using \emph{onset of event} definition across a range of thresholds, we found that non-trivial scaling behaviour and significant group differences remained largely preserved. However, this representation did not produce robust dynamic scaling during the first steps of PRG, as the resulting binary variables are intrinsically sparse at the microscopic level. A comparison of the resulting binary representations is provided in Supplemental Material (Fig.~S1). We additionally tested a \emph{suprathreshold} binarization approach, which generates even sparser binary activity for BOLD signals. In this case, scaling behaviour was generally less robust and failed across most threshold values. Together, these observations indicate that alternative event definitions primarily affect the dynamic scaling results, whereas the static scaling properties remain comparatively robust.

Because the primary binarization retains all suprathreshold samples, we additionally tested whether the observed group differences could be explained by differences in threshold occupancy or the duration of suprathreshold excursions. Event rate, defined as the proportion of suprathreshold ROI--time samples, did not differ significantly between controls and patients across
the examined thresholds (Fig.~\ref{cross-val}B). Similarly, mean event duration did not differ significantly between groups at any threshold (Fig.~\ref{cross-val}C). Thus, the observed differences in PRG scaling exponents cannot be attributed to systematic group differences in the overall density or duration of suprathreshold events.

\subsection*{Alterations in Global Signal Dynamics: Evidence from PSD and DFA}

There is growing evidence that abnormalities in psychosis involve both local and global components of brain activity that are at least partially independent from one another \cite{yang2014altered}. Global signal is average time series computed over all voxels within the brain, and, importantly, its fluctuations in resting-state fMRI have been shown to reflect underlying neural activity fluctuations in the cerebral cortex, supporting their interpretation as signatures of large-scale collective dynamics \cite{scholvinck2010neural}. Moreover, because the global signal reflects the degree of spatial homogeneity in brain activity at each time point, it is particularly sensitive to spatially distributed fluctuations across the brain \cite{liu2017global}. In the context of our PRG results, which indicate altered collective organization under coarse-graining, these findings motivate the investigation of temporal scaling properties directly in the global fMRI signal using PSD and DFA.

\begin{figure}[h!]
\begin{center}
\includegraphics[width=\columnwidth]{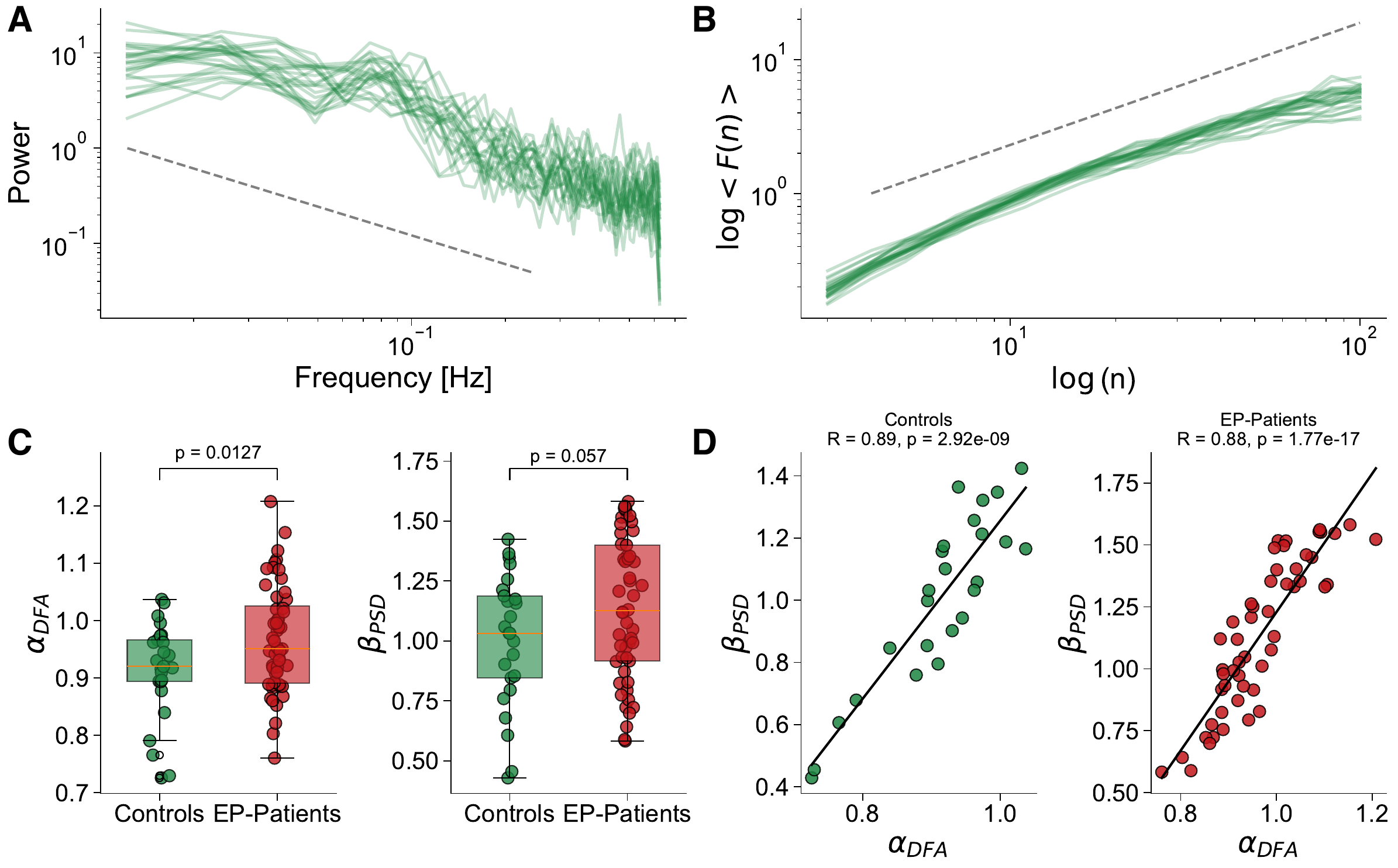}
\caption{
\textbf{Temporal scaling of global signals.}
\textbf{A.} Power spectra of global signals from individual control subjects shown in log-log scale. Each line corresponds to one subject. The dashed gray line indicates the frequency range used for estimating the scaling exponent $\beta_{PSD}$. \textbf{B.} DFA plots of global signals from control subjects, showing the fluctuation function $F(n)$ measured at different time-window lengths and plotted in double-logarithmic scale. Each line corresponds to one subject. The dashed gray line indicates the fitting range used to estimate the scaling exponent $\alpha_{DFA}$. \textbf{C.} Group differences. Boxplots display the median and interquartile range, with individual subjects overlaid. Statistical significance was assessed using Welch's $t$-test. \textbf{D.} DFA scaling exponents correlate with PSD scaling exponents in controls (left) and patients (right). Each point represents one subject. Solid black lines indicate linear regression fits, and titles report Pearson correlation coefficients and associated $p$-values.}
\label{temporal-scaling}
\end{center}
\end{figure}

Fig.~\ref{temporal-scaling} shows the temporal scaling properties of the global fMRI signal in controls and patients assessed using PSD and DFA. The PSD curves exhibited power-law scaling over the selected low-frequency range (Fig.~\ref{temporal-scaling}A), while DFA revealed long-range temporal correlations characterized by scale-invariant growth of the fluctuation function across time windows (Fig.~\ref{temporal-scaling}B). At the group level, patients showed significantly larger DFA exponents compared to controls ($p = 0.0127$), indicating stronger long-range temporal correlations in the global signal (Fig.~\ref{temporal-scaling}C, left). PSD exponents also tended to increase in patients($p = 0.057$; Fig.~\ref{temporal-scaling}C, right). Importantly, DFA and PSD exponents were strongly correlated across subjects in both controls ($R = 0.89$, $p = 2.92 \times 10^{-9}$) and patients ($R = 0.88$, $p = 1.77 \times 10^{-17}$). Together, these findings indicate that large-scale spontaneous fluctuations in psychosis exhibit enhanced temporal persistence and altered scale-free organization. In light of the PRG results, the increase in temporal scaling exponents suggests that the altered collective dynamics observed under coarse-graining are accompanied by changes in the temporal organization of global brain activity.

To assess whether the observed group differences could be accounted for by demographic differences between groups, we fitted OLS models including diagnosis, age, and sex as predictors for each PRG and conventional temporal scaling exponent. The effect of diagnosis remained significant for all six scaling measures after adjustment for age and sex (Table~S2-B). Age was additionally associated with $\mu$ and $z$, whereas no significant effects of sex were observed. Given the group difference in mean framewise displacement, we also tested whether head motion accounted for the temporal scaling differences. Diagnosis remained a significant predictor of $z$, $\alpha_{DFA}$, and $\beta_{PSD}$ after adjustment for mean framewise displacement. Finally, because all patients were receiving antipsychotic medication, we examined medication exposure within the patient group. Lifetime antipsychotic exposure was not significantly associated with any of the four PRG exponents after adjustment for age and sex (all $p>0.13$).

\section*{Discussion}

Schizophrenia is a complex and highly heterogeneous neuropsychiatric disorder whose underlying mechanisms remain incompletely understood. Progress in understanding the disorder has been complicated not only by variability across patient populations, but also by the challenge of integrating findings across different imaging modalities and methodological frameworks. In this context, our work combined coarse graining analyses based on PRG method with temporal scaling measures derived from PSD and DFA to characterize large-scale collective brain dynamics. 

Our findings show non-trivial scaling behavior in both healthy controls and individuals with early psychosis, accompanied by systematic and reproducible shifts in scaling exponents across both multiscale and temporal measures. Importantly, the persistence of non-trivial scaling should not be interpreted as evidence that either group operates precisely at criticality, as model systems can retain scaling away from the critical point while the corresponding exponents change gradually as the system is tuned away from criticality \cite{nicoletti2020scaling}. Instead, the observed exponent shifts are more appropriately considered in terms of changes in scaling organization and their possible relation to relative proximity to criticality.

These results support the view that large-scale brain activity occupies a multidimensional scaling regime within which collective dynamics can be reorganized while preserving broad scale-invariant structure. In the present data, early psychosis did not abolish scaling behavior. Rather, different observables shifted in distinct directions, indicating that collective organization and temporal persistence may be differentially affected. This interpretation is consistent with theoretical perspectives suggesting that heterogeneous neural systems can exhibit extended critical-like regimes in which scale-invariant behavior persists while scaling properties vary depending on network structure and dynamical conditions \cite{moretti2013griffiths}. Within this broader framework, the coexistence of preserved scaling behavior with systematic exponent shifts may reflect a reorganization within a near-critical dynamical regime rather than a simple transition away from criticality.

The observed exponent shifts further suggest that different scaling observables capture partially distinct aspects of collective behaviour. In patients, the increase in the silence probability exponent $\beta$ together with the reduction in the variance scaling exponent $\alpha$ is consistent with weaker large-scale coordination under coarse-graining, reflecting a reduced degree of collective integration across functional scales. In contrast, the increase in the dynamical exponent $z$ and the elevated DFA exponents indicate stronger temporal persistence and enhanced long-range temporal correlations. Rather than reflecting a uniform shift toward either a more ordered or more disordered dynamical state, this dissociation suggests a redistribution between collective integration and temporal persistence in psychosis. Within this framework, psychosis may involve reduced efficiency of large-scale coordination together with increased persistence or rigidity of ongoing activity patterns. Such an interpretation is broadly consistent with previous work demonstrating altered large-scale integration and spatiotemporal organization of resting-state activity in schizophrenia spectrum disorders \cite{garrity2007aberrant, bassett2012altered}. From a criticality perspective, this dissociation may also indicate that spatial and temporal observables reflect distinct changes in proximity to criticality. More direct estimates of proximity to criticality could provide a complementary perspective on this question. Recent approaches have proposed model-based measures of distance to criticality based on spatial covariance structure \cite{dahmen2019second} and temporal dynamics \cite{sooter2025defining}, whose application to fMRI represents an important direction for future work. Such approaches may help determine whether the spatial and temporal alterations observed here reflect distinct changes in proximity to critical regimes.

The observed alterations in scaling organization may also be interpretable in light of leading mechanistic theories of schizophrenia and psychosis. A large body of work implicates disruptions in excitation–inhibition (E/I) balance, particularly involving NMDA receptor hypofunction on parvalbumin-positive inhibitory interneurons, as a central mechanism underlying abnormal cortical dynamics in schizophrenia \cite{olney1995glutamate, lisman2008circuit, krystal2017impaired}. Within theoretical and computational models of large-scale brain dynamics, E/I balance is thought to play a central role in regulating proximity to critical regimes \cite{shew2013functional, poil2012critical, munoz2018colloquium}. From this perspective, the altered scaling exponents observed in early psychosis may reflect changes in the balance between collective integration and temporal persistence arising from dysregulated cortical gain, recurrent excitation, and inhibitory control. 

In particular, NMDA-receptor hypofunction has been proposed to weaken the stability of coordinated large-scale network interactions while simultaneously increasing internally generated activity fluctuations and aberrant persistence of neural states \cite{durstewitz2008dual, krystal2017impaired}. Such mechanisms could qualitatively account for the present combination of weaker collective integration under coarse-graining together with enhanced temporal persistence observed in psychosis. More broadly, contemporary theories of schizophrenia increasingly emphasize abnormalities in large-scale integration, dynamical flexibility, and hierarchical coordination across cortical systems, including dysconnectivity between distributed brain networks \cite{friston1998disconnection, stephan2009dysconnection}, impaired context-dependent gain modulation \cite{phillips2003convergence, adams2013computational}, and excessive stability or rigidity of internally generated representations \cite{rolls2008computational}. Within this broader framework, the altered scaling relations observed here may reflect a systems-level signature of disturbed collective dynamics arising from underlying circuit-level dysregulation. Importantly, the present findings do not permit direct inference about specific neurobiological mechanisms. The observed exponent shifts should therefore be interpreted as markers of altered large-scale brain dynamics rather than direct evidence for any particular pathophysiological process. Nevertheless, they suggest that scaling observables may provide a quantitative bridge between microscale circuit theories of psychosis and macroscale alterations in whole-brain dynamics. Future work combining criticality-inspired analyses with computational models of E/I balance, pharmacological manipulations of NMDA signaling, and multimodal electrophysiological recordings may help clarify how specific circuit mechanisms contribute to altered large-scale dynamics in psychiatric disorders.

Although deviations from criticality-related measures have previously been reported in psychosis spectrum disorders \cite{fekete2021multiscale, lee2021alteration, alamian2022altered, dick2022fractal}, the present study represents, to our knowledge, the first application of a phenomenological renormalization group (PRG) framework to a clinical population. The PRG approach offers several advantages for studying large-scale brain organization. Unlike connectivity-based approaches that impose predefined network architectures or fixed spatial neighborhoods, PRG directly probes how correlated activity reorganizes under iterative coarse-graining, allowing collective dynamics to emerge from the empirical correlation structure itself. Importantly, the framework does not require specifying a particular microscopic model of neural interactions or constraining the analysis using an assumed structural connectivity architecture. In this context, PRG provides a principled framework for characterizing scaling organization directly from functional activity while remaining relatively agnostic about the underlying mechanisms generating the observed dynamics.

A few important limitations of the present study should be considered. First, fMRI provides only an indirect measure of neural activity through slow hemodynamic fluctuations, limiting the temporal precision with which critical dynamics can be characterized. Importantly, scaling exponents derived from criticality-inspired analyses can depend on event definitions and representational choices, although the principal findings remained robust across multiple thresholds and binarization schemes. The relatively limited duration and low temporal resolution of fMRI recordings make standard neuronal avalanche analyses, originally developed for electrophysiological data, difficult to apply reliably. Second, all patients included in the present study were receiving medication, making it difficult to disentangle disease-related alterations from potential medication effects on large-scale brain dynamics. Finally, although the observed scaling exponents differentiated groups, we did not observe strong associations with clinical symptom measures within the present cohort. One possibility is that these scaling properties reflect relatively stable organizational features of large-scale brain dynamics rather than momentary symptom fluctuations. Alternatively, the relatively early disease stage and limited symptom variability of the current sample may have reduced sensitivity to clinical associations. This interpretation is broadly consistent with recent work reporting stronger and more robust functional connectivity–symptom relationships in chronic psychosis compared to early psychosis populations \cite{foster2025connectome}. Future longitudinal studies will be important for determining whether scaling observables track disease progression, treatment response, or cognitive dysfunction over time.

In summary, our findings demonstrate that resting-state brain activity in early psychosis retains broad signatures of scale-invariant organization while exhibiting systematic alterations in spatiotemporal scaling relations relative to healthy controls. By complementing PRG analysis with DFA and PSD approaches, this work provides a multiscale framework for probing collective brain dynamics in psychiatric disorders. More generally, the results support the idea that psychiatric disorders may involve altered organization of large-scale dynamical states rather than isolated disruptions of specific neural processes, and suggest that scaling measures may serve as useful quantitative markers in both fundamental and clinical research.

\subsection*{Data and Code availability}
We used a publicly available fMRI dataset from Human Connectome Project (HCP) which is available at (\url{https://www.humanconnectome.org/study/human-connectome-project-for-early-psychosis}). Codes to perform PRG analysis are publicly available on GitHub at \url{https://github.com/lffrrnt/CoarseGraining}.

\selectlanguage{english}
\FloatBarrier
\bibliographystyle{jneurosci}
\bibliography{bibliography/biblio}

\clearpage

\section*{Supplementary Materials}

\renewcommand{\thefigure}{S\arabic{figure}}
\setcounter{figure}{0}

\renewcommand{\thetable}{S\arabic{table}}
\setcounter{table}{0}

\begin{table}[htbp]
\centering
\caption{\textbf{Demographics.} Data are presented as the means $\pm$ standard deviation (SD). 
P-values for age were calculated using the Kruskal-Wallis H test, while p-values for gender 
were obtained using a two-tailed Pearson chi-square test.}
\label{tab:demographics}

\begin{tabular}{lccc}
\hline
 & Patient (n=52) & Controls (n=25) & p-value \\
\hline
Gender (F/M) & 16/36 & 11/14 & 0.0171 \\
Age (years) & $22.48 \pm 3.43$ & $23.91 \pm 3.76$ & 0.0846 \\
Average Medication Exposure (months) & $17.75 \pm 15.24$ & N/A & -- \\
Positive & $11.27 \pm 3.37$ & N/A & -- \\
Negative & $16.12 \pm 6.06$ & N/A & -- \\
General Psychopathology & $26.38 \pm 4.58$ & N/A & -- \\
Total & $53.77 \pm 9.61$ & N/A & -- \\
\hline
\end{tabular}
\end{table}

\subsection*{Supplemental Text S1. Image Pre-processing with fMRIPrep}
Results included in this manuscript come from preprocessing performed using fMRIPrep 24.1 \cite{esteban2019fmriprep}, which is based on Nipype 1.6.1 \cite{gorgolewski2011}. The T1-weighted (T1w) image was corrected for intensity non-uniformity (INU) with \texttt{N4BiasFieldCorrection} \cite{tustison2010n4itk}, distributed with ANTs 2.3.3 \cite{avants2008symmetric}, and used as T1w-reference throughout the workflow for each participant. The
T1w-reference was then skull-stripped with a Nipype implementation of the
\texttt{antsBrainExtraction.sh} workflow (from ANTs), using OASIS30ANTs as the target template.
Brain tissue segmentation of cerebrospinal fluid (CSF), white-matter (WM) and grey-matter
(GM) was performed on the brain-extracted T1w using \texttt{fast} (FSL 6.0.5.1:57b01774) \cite{zhang2002segmentation}. Brain surfaces were reconstructed using \texttt{recon-all} (FreeSurfer 6.0.1) \cite{dale1999cortical}, and the brain mask estimated previously was refined with a custom variation of the
method to reconcile ANTs-derived and FreeSurfer-derived segmentations of the cortical grey
matter of Mindboggle \cite{klein2017mindboggling}. Volume-based spatial normalisation to one
standard space (MNI152NLin2009cAsym) was performed through nonlinear registration with
\texttt{antsRegistration} (ANTs 2.3.3), using brain-extracted versions of both T1w reference and the
T1w template.
For each of the BOLD runs per subject, the following preprocessing was performed: First, a
reference volume and its skull-stripped version were generated using a custom methodology
of fMRIPrep. Head-motion parameters with respect to the BOLD reference (transformation
matrices and six corresponding rotation and translation parameters) are estimated before any
spatiotemporal filtering using \texttt{mcflirt} (FSL 6.0.5.1:57b01774) \cite{jenkinson2002improved}.
BOLD runs were slice-time corrected to 1.46s (0.5 of the slice acquisition range 0s--2.92s)
using \texttt{3dTshift} from AFNI \cite{cox1997software}. The BOLD time-series (including
slice-timing correction when applied) were resampled onto their original, native space by
applying the transforms to correct for head-motion. These resampled BOLD time-series will
be referred to as preprocessed BOLD in original space, or just preprocessed BOLD. The
BOLD reference was then co-registered to the T1w reference using \texttt{bbregister} (FreeSurfer),
which implements boundary-based registration \cite{greve2009accurate}. Co-registration was
configured with six degrees of freedom.
Several confounding time-series were calculated
based on the preprocessed BOLD: framewise displacement (FD), DVARS, and three
region-wise global signals. FD was computed using two formulations following Power
(absolute sum of relative motions) \cite{power2014methods} and Jenkinson (relative root mean
square displacement between affines) \cite{jenkinson2002improved}. FD and DVARS are calculated
for each functional run, both using their implementations in Nipype (following the definitions
by \cite{power2014methods}). The three global signals were extracted within the CSF, the WM, and
the whole-brain masks.
Additionally, a set of physiological regressors was extracted to allow
for component-based noise correction (CompCor) \cite{behzadi2007component}. Principal
components were estimated after high-pass filtering the preprocessed BOLD time-series
(using a discrete cosine filter with 128s cut-off) for the two CompCor variants: temporal
(tCompCor) and anatomical (aCompCor). tCompCor components are then calculated from
the top $2\%$ variable voxels within the brain mask. For aCompCor, three probabilistic masks
(CSF, WM, and combined CSF+WM) are generated in anatomical space. The implementation
differs from that of \cite{behzadi2007component} in that, instead of eroding the masks by 2 pixels on
BOLD space, the aCompCor masks are subtracted from a mask of pixels that likely contain a
volume fraction of GM. This mask is obtained by dilating a GM mask extracted from the
FreeSurfer's \texttt{aseg} segmentation, and it ensures components are not extracted from voxels
containing a minimal fraction of GM.
Finally, these masks are resampled into BOLD space
and binarized by thresholding at 0.99 (as in the original implementation). Components are
also calculated separately within the WM and CSF masks. For each CompCor decomposition,
the $k$ components with the largest singular values are retained, such that the retained
components' time series are sufficient to explain $50 \%$ of variance across the nuisance
mask (CSF, WM, combined, or temporal). The remaining components are dropped from
consideration.
The head-motion estimates calculated in the correction step were also placed
within the corresponding confounds file. The confound time series derived from head motion
estimates and global signals were expanded with the inclusion of temporal derivatives and
quadratic terms for each \cite{satterthwaite2013improved}. Frames that exceeded a threshold of 0.5
mm FD or 1.5 standardised DVARS were annotated as motion outliers.
All resamplings can be performed with a single interpolation step by composing all the pertinent transformations
(i.e.\ head-motion transform matrices, susceptibility distortion correction when available, and
co-registrations to anatomical and output spaces). Gridded (volumetric) resamplings were
performed using \texttt{antsApplyTransforms} (ANTs), configured with Lanczos interpolation to
minimise the smoothing effects of other kernels \cite{lanczos1964evaluation}. Non-gridded (surface)
resamplings were performed using \texttt{mri\_vol2surf} (FreeSurfer).
Many internal operations of fMRIPrep use Nilearn 0.8.1 \cite{abraham2014machine}, mostly
within the functional processing workflow. For more details, see the fMRIPrep website
(\url{https://fmriprep.org}).

\begin{figure}
\begin{center}
\includegraphics[width=\columnwidth]{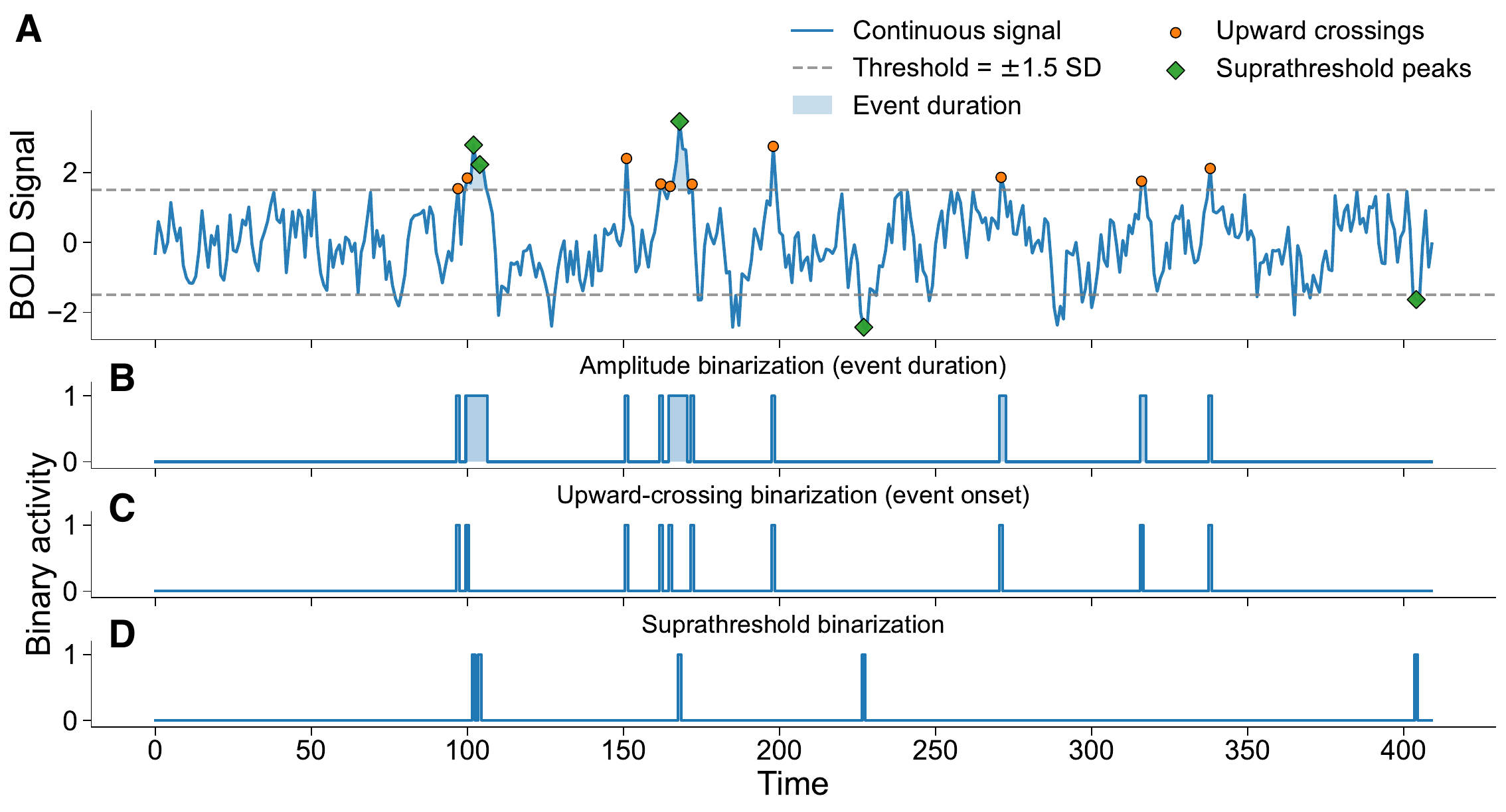}
\caption{\textbf{Comparison of event detection methods.} \textbf{A.} Example BOLD time series with a threshold set at $\pm 1.5$ SD (grey dashed lines). Orange markers indicate upward threshold crossings, green diamonds denote suprathreshold peaks, and shaded regions represent time points exceeding the threshold. \textbf{B.} Amplitude-based binarization (event duration), where all time points above the threshold are assigned a value of $1$. \textbf{C.} Upward-crossing binarization (event onset), where only upward threshold crossings are identified as discrete events. \textbf{D.} Suprathreshold binarization, where only local maxima and minima within suprathreshold excursions are retained as discrete events.}
\label{binarization}
\end{center}
\end{figure}

\begin{figure}
\centering

\includegraphics[width=\columnwidth]{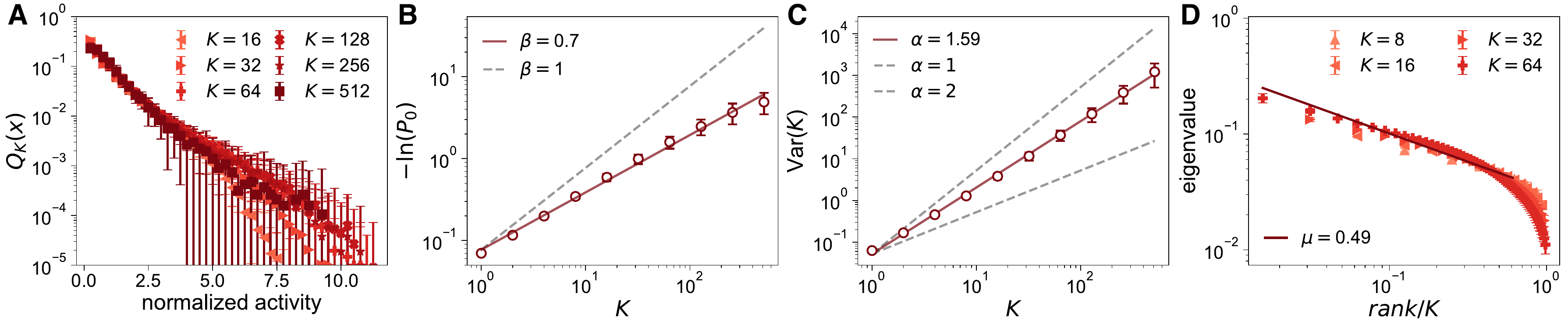}

\vspace{0.3cm}

\includegraphics[width=.5\columnwidth]{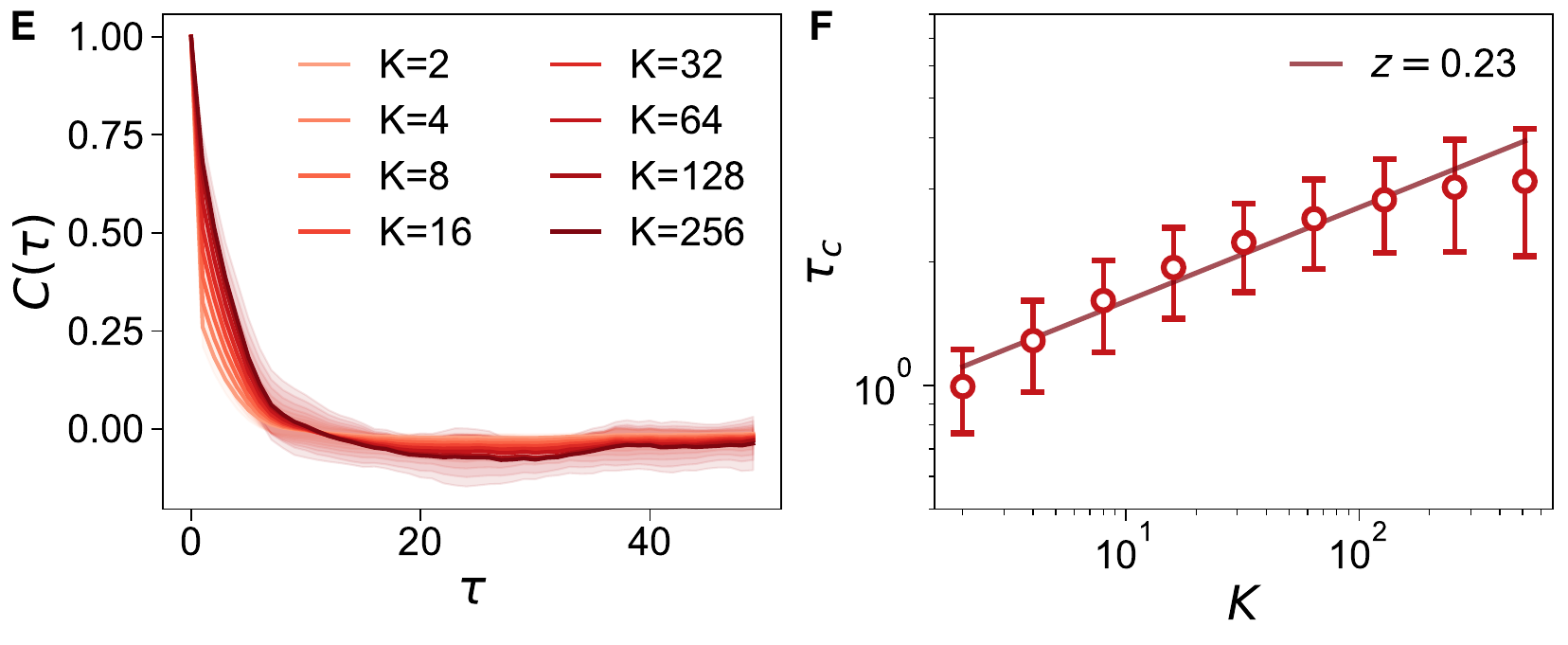}
\caption{\textbf{Scaling relations disappear when cross-correlations are destroyed.} \textbf{A.} Average probability distribution of normalized non-zero coarse-grained activity across 25 phase-shuffled subjects. In contrast to the empirical data, the activity distribution approaches a Gaussian form for large cluster sizes. \textbf{B.} Average probability of silence in coarse-grained variables, expressed as $-\ln P_0$, as a function of cluster size $K$. The probability of silence follows $P_0(K)=e^{-bK^\beta}$, with $\beta=1$ corresponding to the expectation for independent variables ($\beta=0.94$, $R^2=0.99$, $p < 10^{-9}$). \textbf{C.} Average variance of the non-normalized coarse-grained variables as a function of cluster size $K$. Solid lines indicate linear least-squares fits in double logarithmic scale according to $\log \mathrm{Var}(K)=\alpha \log K + b$, where $\alpha=1$ is expected for independent variables ($\alpha=1.03$, $R^2=0.99$, $p < 10^{-9}$). \textbf{D.} Eigenvalue spectra of covariance matrices within clusters for different cluster sizes. In the randomized data, the data collapse observed in the empirical data is lost. Error bars represent standard deviation across subjects (SD). \textbf{E.} Average autocorrelation functions $C(\tau)$ for different cluster sizes $K$. For each subject, autocorrelations were computed and averaged across clusters, and the displayed curves correspond to the mean across subjects. Shaded regions represent the standard deviation across subjects. \textbf{F.} Autocorrelation times $\tau_c$, obtained by fitting exponential decays to the subject-specific average autocorrelation functions. Points indicate the mean $\tau_c$ across subjects as a function of cluster size $K$, with error bars denoting standard deviation. Unlike the empirical data, the phase-shuffled signals do not exhibit a power-law dependence of $\tau_c$ on cluster size, indicating the absence of robust dynamic scaling after the destruction of cross correlations.}
\label{controls-dy-surrogates}
\end{figure}

\begin{figure}
\centering

\includegraphics[width=\columnwidth]{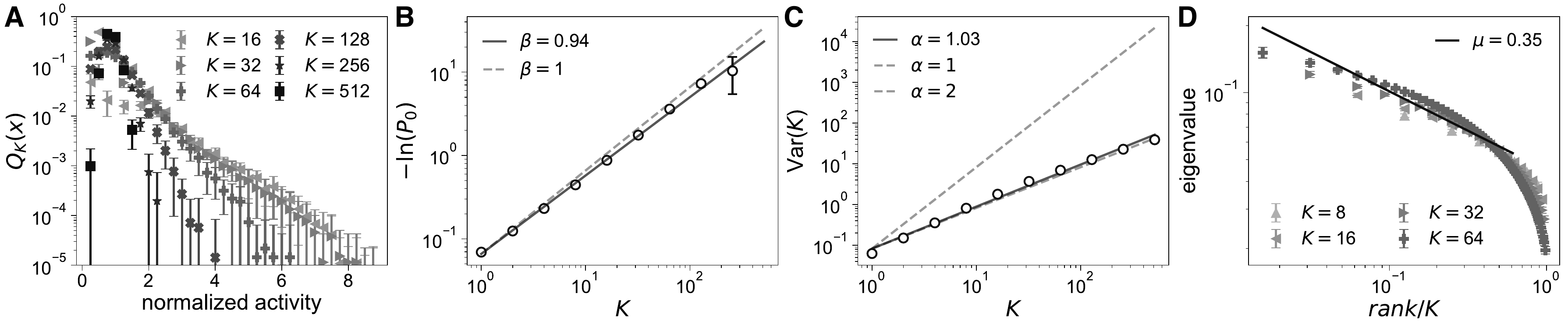}

\vspace{0.3cm}

\includegraphics[width=.5\columnwidth]{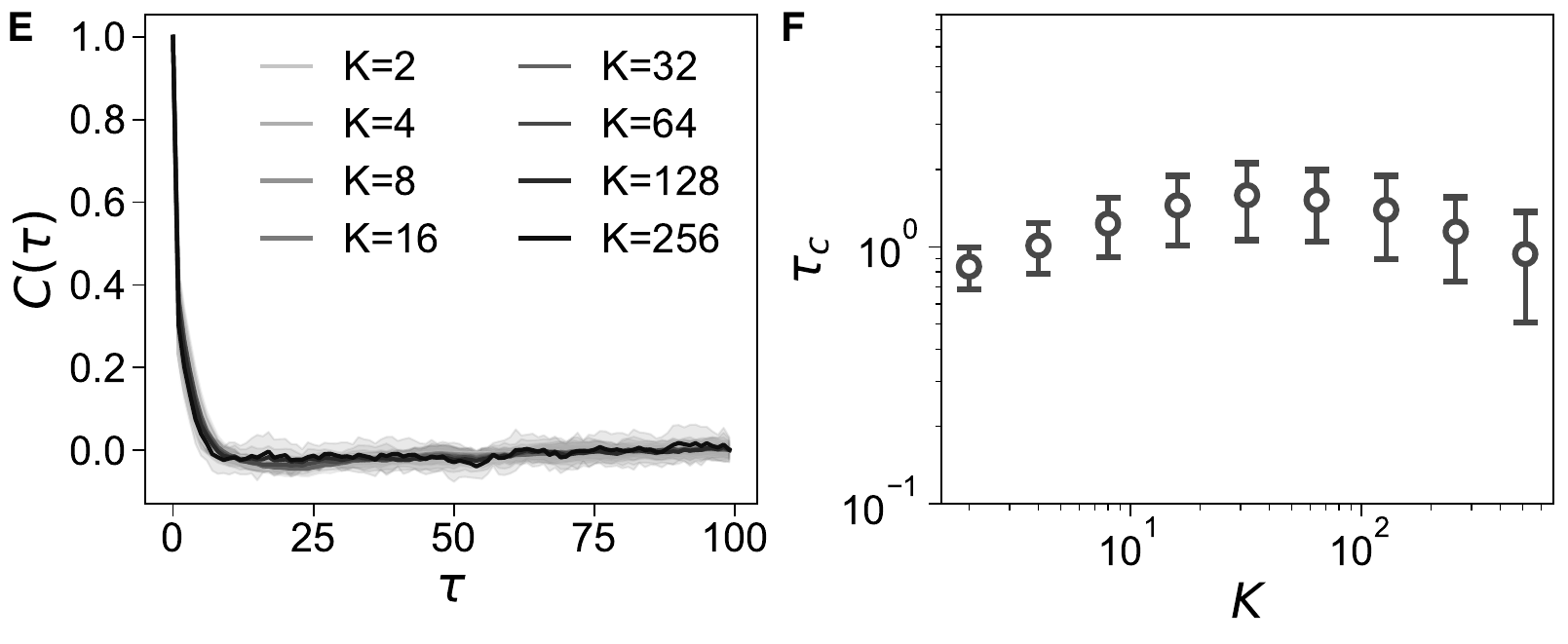}

\caption{\textbf{Resting-state whole-brain activity in early psychosis exhibits altered statistical signatures of criticality.}
\textbf{A.} Average probability distribution of normalized nonzero activity across 52 early-psychosis subjects. Darker red shades correspond to larger cluster sizes.
\textbf{B.} Average log-probability of silence in coarse-grained variables ($-\ln P_0$) as a function of cluster size $K$. The solid line indicates a linear least-squares fit of the scaling relation $-\ln P_0 \propto K^\beta$, yielding $\beta = 0.7$ ($R^2=0.99$, $p < 10^{-9}$). The gray dashed line denotes the expected behavior for uncorrelated variables ($\beta = 1$).
\textbf{C.} Average variance of the non-normalized coarse-grained variables as a function of cluster size $K$. The solid line corresponds to a linear least-squares fit in log--log space according to $\log \mathrm{Var}(K)=\alpha \log K + b$, with $\alpha = 1.59$ ($R^2=0.99$, $p < 10^{-9}$). Gray dashed lines indicate reference scaling behaviors for uncorrelated ($\alpha = 1$) and fully correlated ($\alpha = 2$) variables.
\textbf{D.} Eigenvalue spectra of covariance matrices for clusters with size $K \leq 64$. The solid line represents a least-squares fit of the relation $\log \lambda = -\mu \log(rank/K) + \log b$, performed for $K = 64$. Error bars denote standard deviation across patients (SD).
\textbf{E.} Average autocorrelation functions $C(\tau)$ for each cluster size $K$.
\textbf{F.} Autocorrelation time $\tau_c$, obtained by fitting an exponential decay to the average autocorrelation function of each patient individually. $z=0.22 \pm 0.05$ (n=52). The points show the mean $\tau_c$ across patients as a function of $K$, with error bars denoting the 
standard deviation. A power-law scaling $\tau_c \propto K^z$ is observed; the reported exponent $z=0.23$ is obtained from a least-squares fit to the averaged $\tau_c$ values ($R^2=0.96$, $p < 10^{-4}$).}

\label{fig:patients-dy}
\end{figure}

\begin{figure}
\begin{center}
\includegraphics[width=\columnwidth]{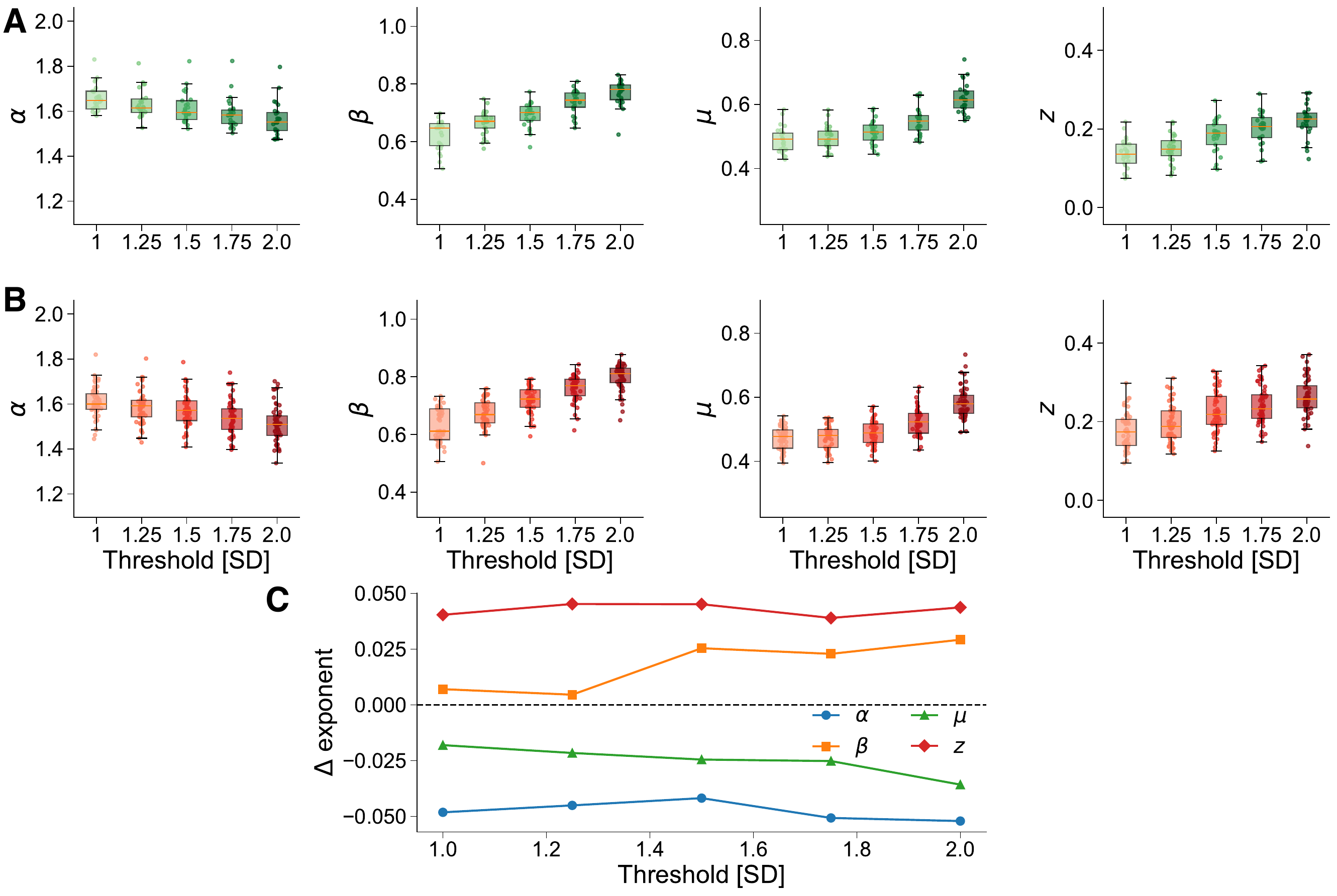}
\caption{\textbf{Robustness of PRG exponents with respect to binarization threshold.}
\textbf{A.} Distribution of PRG exponents in healthy controls for five different binarization thresholds ranging from $1$ to $2$ SD. Box plots summarize the subject distributions for the variance exponent $\alpha$, silence probability exponent $\beta$, eigenspectrum exponent $\mu$, and dynamical exponent $z$. \textbf{B.} Same analysis for early-psychosis patients. Across both groups, the estimated exponents exhibit relatively limited variability as a function of threshold choice. 
\textbf{C.} Mean group differences in PRG exponents as a function of binarization threshold, computed as 
$\Delta \xi = \langle \xi_{\mathrm{patients}} \rangle - \langle \xi_{\mathrm{controls}} \rangle$, 
where $\xi \in \{\alpha,\beta,\mu,z\}$. The direction of the group differences remains consistent across thresholds, indicating that the observed alterations in scaling behavior are robust with respect to the choice of binarization threshold.}
\label{threshold}
\end{center}
\end{figure}

\begin{table}
\centering
\caption{
\textbf{Additional robustness analyses of group differences.}
\textbf{A.} Robustness of PRG group differences across parcellation resolutions. Reported values are Welch's $t$-test $p$-values for comparisons between healthy controls and early psychosis participants.
\textbf{B.} Ordinary least squares (OLS) regression analyses including
diagnosis, age, and sex as predictors. Diagnosis was coded with healthy
controls as the reference group. The diagnosis coefficient
$\beta_{\mathrm{OLS}}$ represents the adjusted effect of early psychosis
relative to healthy controls.}
\label{tab:robustness}

\vspace{0.5em}

\textbf{A. Parcellation resolution}

\vspace{0.3em}

\begin{tabular}{lcccc}
\hline
\textbf{Number of ROIs} & $\beta$ & $\alpha$ & $\mu$ & $z$ \\
\hline
800 & 0.116   & 0.00439 & 0.0211  & $1.41\times10^{-5}$ \\
600 & 0.0199  & 0.00329 & 0.0131  & 0.000205 \\
400 & 0.00613 & 0.0112  & 0.00969 & $1.36\times10^{-5}$ \\
\hline
\end{tabular}

\vspace{1.2em}

\textbf{B. Adjustment for age and sex}

\vspace{0.3em}

\begin{tabular}{lccccc}
\hline
\textbf{Exponent} &
\textbf{Diagnosis $\beta_{\mathrm{OLS}}$} &
\textbf{95\% CI} &
\textbf{Diagnosis $p$} &
\textbf{Age $p$} &
\textbf{Sex $p$} \\
\hline
$\alpha$ &
$-0.045$ & $[-0.082,\,-0.008]$ & 0.018 & 0.286 & 0.879 \\

$\beta$ &
\phantom{$-$}0.024 & $[0.001,\,0.046]$ & 0.038 & 0.724 & 0.599 \\

$\mu$ &
\phantom{$-$}0.030 & $[0.012,\,0.048]$ & 0.002 & 0.001 & 0.737 \\

$z$ &
\phantom{$-$}0.050 & $[0.026,\,0.073]$ & $<0.001$ & 0.029 & 0.707 \\

$\alpha_{\mathrm{DFA}}$ &
\phantom{$-$}0.053 & $[0.007,\,0.099]$ & 0.024 & 0.653 & 0.923 \\

$\beta_{\mathrm{PSD}}$ &
\phantom{$-$}0.147 & $[0.000,\,0.294]$ & 0.049 & 0.609 & 0.533 \\
\hline
\end{tabular}

\end{table}

\end{CJK}
\end{document}